\documentclass{elsarticle}

\usepackage{amssymb,amsmath}
\usepackage{graphics}
\usepackage[caption=false]{subfig}
\usepackage{hyperref}

\usepackage[usenames,dvipsnames]{color}
\usepackage{ulem}

\graphicspath{{.},{../Figs/}} 

\begin{document}

\sloppy

\begin{frontmatter}
\title{An improved dissipative coupling scheme for a system of 
Molecular Dynamics particles interacting with a Lattice Boltzmann 
fluid}
\author[mpip]{Nikita Tretyakov\corref{mycorrespondingauthor}}
\cortext[mycorrespondingauthor]{Corresponding author}
\ead{tretyakov@mpip-mainz.mpg.de}
\author[mpip,monash]{Burkhard D\"unweg}
\address[mpip]{Max-Planck-Institut f\"{u}r Polymerforschung, 
  Ackermannweg 10, 55128 Mainz, Germany}
\address[monash]{Department of Chemical Engineering, Monash University,
  Clayton, Victoria 3800, Australia}

\begin{abstract}
  We consider the dissipative coupling between a stochastic Lattice
  Boltzmann (LB) fluid and a particle-based Molecular Dynamics (MD)
  system, as it was first introduced by Ahlrichs and D\"unweg
  (J. Chem. Phys. 111 (1999) 8225). The fluid velocity at the position
  of a particle is determined by interpolation, such that a Stokes
  friction force gives rise to an exchange of momentum between the
  particle and the surrounding fluid nodes. For efficiency reasons,
  the LB time step is chosen as a multiple of the MD time step, such
  that the MD system is updated more frequently than the LB fluid. In
  this situation, there are different ways to implement the coupling:
  Either the fluid velocity at the surrounding nodes is only updated
  every LB time step, or it is updated every MD step. It is
  demonstrated that the latter choice, which enforces momentum
  conservation on a significantly shorter time scale, is clearly
  superior in terms of temperature stability and accuracy, and
  nevertheless only marginally slower in terms of execution speed. The
  second variant is therefore the recommended implementation.
\end{abstract}


\end{frontmatter}


\section{Introduction}

In the last decades the Lattice Boltzmann (LB)
technique~\cite{Succi01,benzi_lattice_1992,qian_lattice_1992,
  BD_AL_2009,CKA_JRC_2010} has evolved into a well-founded and
efficient numerical tool for the study of fluid mechanics. It has
numerous applications, ranging from the studies of
turbulence~\cite{YH_SSG_LSL_2005} and other macroscopic fluid dynamics
problems~\cite{Succi_2008} to soft matter investigations on the meso-
or microscale. Hydrodynamics of soft matter is in itself a large
field, and LB has been applied to, e.~g., liquid
crystals~\cite{DM_EO_MEC_2007}, two-phase
flows~\cite{MRS_EO_WRO_1996}, binary mixtures~\cite{VMK_MEC_IP_2001},
and hybrid simulations of particle-based systems, like colloids or
polymers, in a solvent. The present paper is a methodological
investigation dealing with this last application, which is based upon
coupling LB to Molecular Dynamics (MD). This method, which will be
referred to by LB/MD, has been described in detail in
Ref.~\cite{BD_AL_2009}.

In colloidal dispersions or polymer solutions the molecular structure
of the solvent is often irrelevant, while dynamic correlations between
the dispersed particles, transmitted via fast momentum transport
through the solvent (the so-called ``hydrodynamic interactions'') are
of paramount importance. There are many ways to take these
correlations into account in a simulation, of which LB/MD is only one.
Competing approaches are Brownian Dynamics
(BD)~\cite{ermak_brownian_1978}, Dissipative Particle Dynamics
(DPD)~\cite{ espanol_statistical_1995}, Multi-Particle Collision
Dynamics (MPCD)~\cite{gompper_multi-particle_2008}, Smoothed
Dissipative Particle Dynamics (SDPD)~\cite{espanol_smoothed_2003}, and
``conventional'' Navier-Stokes equation (NSE)
solvers~\cite{balboausabiaga_staggered_2012}. All these methods have
advantages and disadvantages, and these have (at least partly) been
discussed in Ref.~\cite{BD_AL_2009}. Important criteria that a
simulation method should satisfy are: (i) consistent representation of
thermal fluctuations, which are very important on the small length
scales of soft matter (satisfied by all); (ii) linear scaling
(satisfied by all except BD); and (iii) control over the amplitude of
thermal fluctuations which should depend on the degree of
coarse-graining (or the length-scale resolution) of the simulation
(satisfied only by LB/MD, SDPD, and NSE). LB/MD is particularly
attractive for several reasons: (i) due to the lattice, LB is based on
a tight data structure, which allows efficient memory management; (ii)
due to the streaming-and-collision structure of the algorithm, the
method exhibits a high degree of locality, which makes it amenable to
parallelization based upon geometric domain decomposition. Indeed, in
a comparative study LB/MD was found to be significantly faster than a
DPD simulation of the same physical system~\cite{JS_MS_CH_2009}. A
disadvantage of lattice methods is however their inability to deal
with difficult boundary conditions, in particular in cases where these
involve a deforming simulation cell~\cite{kraynik_extensional_1992}.

It is clear that LB/MD soft-matter simulations have to take care of
two aspects that are foreign to the plain LB method, which is
essentially not much more than an NSE solver: On the one hand, one has
to introduce thermal fluctuations by means of a suitable stochastic
collision operator, and on the other hand, one needs a suitable
coupling scheme for interaction with the particle-based system. The
first aspect has seen significant progress in the last two
decades~\cite{ladd_numerical_1994a,ladd_numerical_1994b,
  RA_KS_MEC_2005,BD_US_AL_2007,BD_US_AL_2009,kaehler_fluctuating_2013},
and this topic shall not be our concern here. For the coupling,
various schemes have been developed. Among the most prominent methods,
one can mention reflecting boundary
conditions~\cite{ladd_numerical_1994a,ladd_numerical_1994b,AL_RV_2001,
  BD_AL_2009}, force coupling~\cite{PA_BD_1999,BD_AL_2009}, the
immersed boundary method (IBM)~\cite{Peskin_2002} and external
boundary forces (EBF)~\cite{JW_CKA_2010}.

The EBF method is applied to extended objects to satisfy a no-slip
boundary condition on their surface. The IBM represents a
fluid-particle interface by a set of Lagrangian nodes and interactions
are applied as body forces to the fluid. These approaches result in a
fairly accurate representation of hydrodynamic boundary conditions,
however at the expense of a somewhat complicated algorithm. On the
other hand, many soft-matter systems involve objects that are quite
large and very ``soft'' (like polymers of various molecular
architectures, tethered membranes, etc.). For these systems the
details of the coupling on the local (or monomer) scale do not matter
very much --- it is only important that the hydrodynamic interactions
are correctly represented on larger scales (larger than the monomer
size but still significantly smaller than the size of the object as a
whole). Therefore, a very simple coupling scheme is desirable, and the
force coupling originally put forward by Ahlrichs and
D\"unweg~\cite{PA_BD_1999} and recently refined by
Schiller~\cite{Schiller_2014} satisfies this criterion. It is also
clear that Ladd's reflecting boundary
method~\cite{ladd_numerical_1994a,ladd_numerical_1994b} is not
suitable for polymer systems, since this would require to model each
monomer as an extended sphere, which would be computationally much
more expensive than the point-particle representation used in force
coupling. It is this latter method upon which we will focus in the
present paper.

The force coupling algorithm is inherently dissipative, i.~e. the
velocity of an MD particle is damped with respect to the velocity of
the LB fluid interpolated to the particle's position. Random forces
are added to the particles to account for thermal noise. It should be
noted that the fluctuation-dissipation theorem stipulates that every
dissipation mechanism needs to be compensated by a corresponding
noise. This means not only that the viscous damping within the LB
fluid must be compensated by a stochastic collision operator, but also
that the damping of the particles relative to the surrounding flow
needs a compensating noise as well. The counterparts of the coupling
forces (damping plus noise) are exerted on the LB fluid to conserve
the total momentum. The calculation of the coupling forces takes place
every MD step, but the LB update typically needs to be done only after
several MD steps. This scheme allows us to capture the dynamics of the
fluid and the immersed particles correctly and reproduce hydrodynamic
behavior. However, the MD and LB timesteps have to be chosen wisely to
find a compromise between the performance and the heat-up of the
particle-based system at moderate friction coefficients, which must be
viewed as a discretization error. It turns out that the details of
this momentum exchange have a significant influence on the size of the
discretization error, and the topic of the present investigation is to
improve the method with respect to this aspect.

A naive and straightforward approach would involve a re-calculation of
the streaming velocities at the surrounding LB nodes only every LB
step. In the present paper, we investigate both this method as well as
a refined one, where the streaming velocities at the surrounding nodes
are rather re-calculated every MD time step, in accord with the
coupling forces. This latter scheme is obviously more accurate, and
gives conservation of total momentum not only on the scale of the LB
time step, but rather of the MD time step. We also find that this
improves the temperature stability of the simulation substantially,
and permits more freedom in the choice of the MD and LB time steps.
Furthermore, the computational overhead associated with the improved
scheme is insignificant, since it employs already available momentum
changes and only adds a few more operations on the surrounding lattice
sites.

The paper is organized as follows: Sec.~\ref{sec:lb_tech} presents a
short overview of the LB method. The details on the coupling technique
and the update scheme are given in Sec.~\ref{sec:coup}, together with
a comparison between the two strategies mentioned above. We conclude
in Sec.~\ref{sec:conc} with a short summary.

\section{Lattice Boltzmann technique}
\label{sec:lb_tech}

In this section we provide a brief explanation of the LB method. For
details and the underlying theory, we refer the reader to the review
given in Ref.~\cite{BD_AL_2009}.

\begin{figure}
  \centering
  \subfloat[]{\label{fig:d3q19}
    \includegraphics[width=0.48\textwidth]{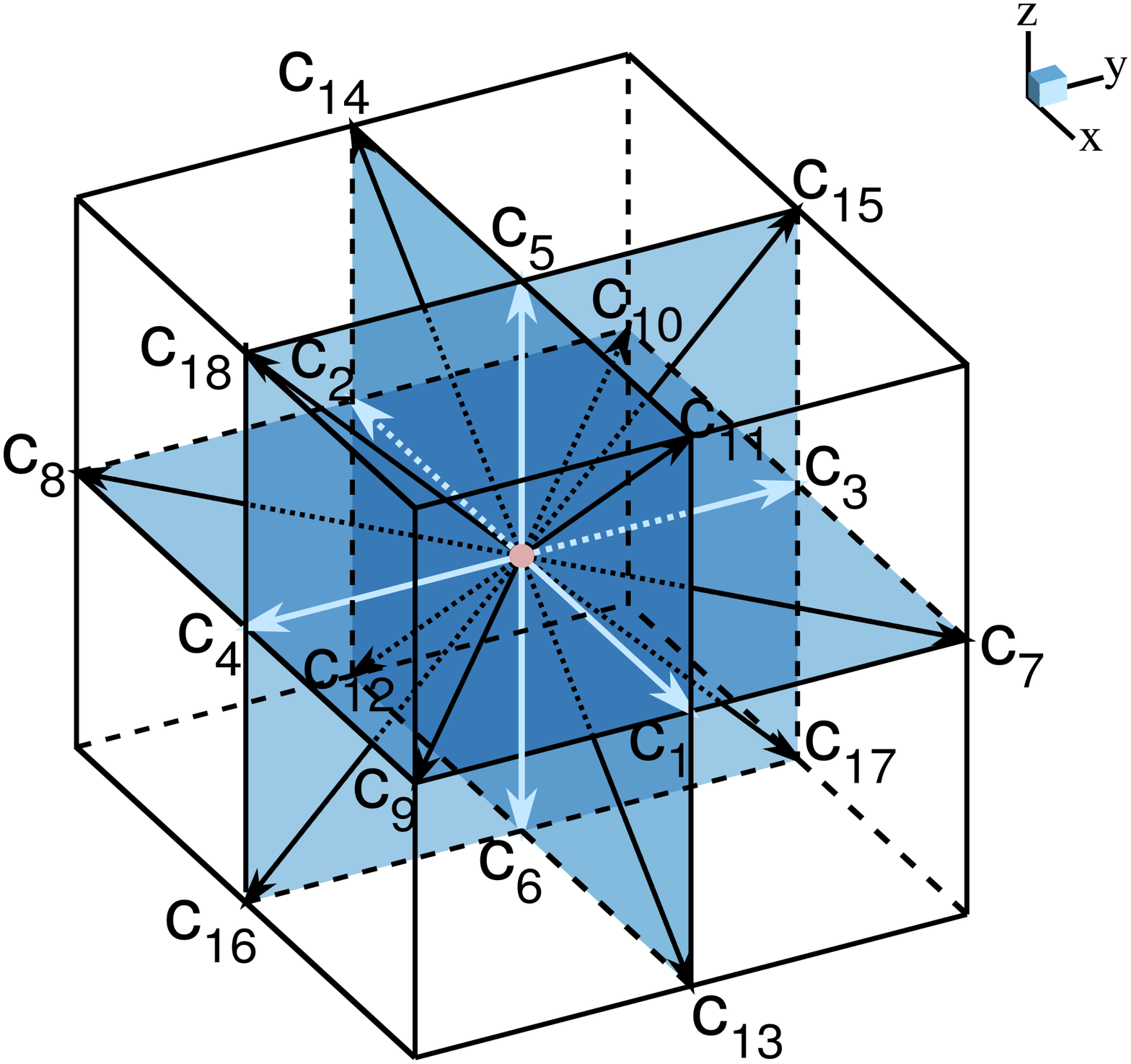}}\hspace*{0.5cm}
  \subfloat[]{\label{fig:LB_MD_coupling}
    \includegraphics[width=0.45\textwidth]{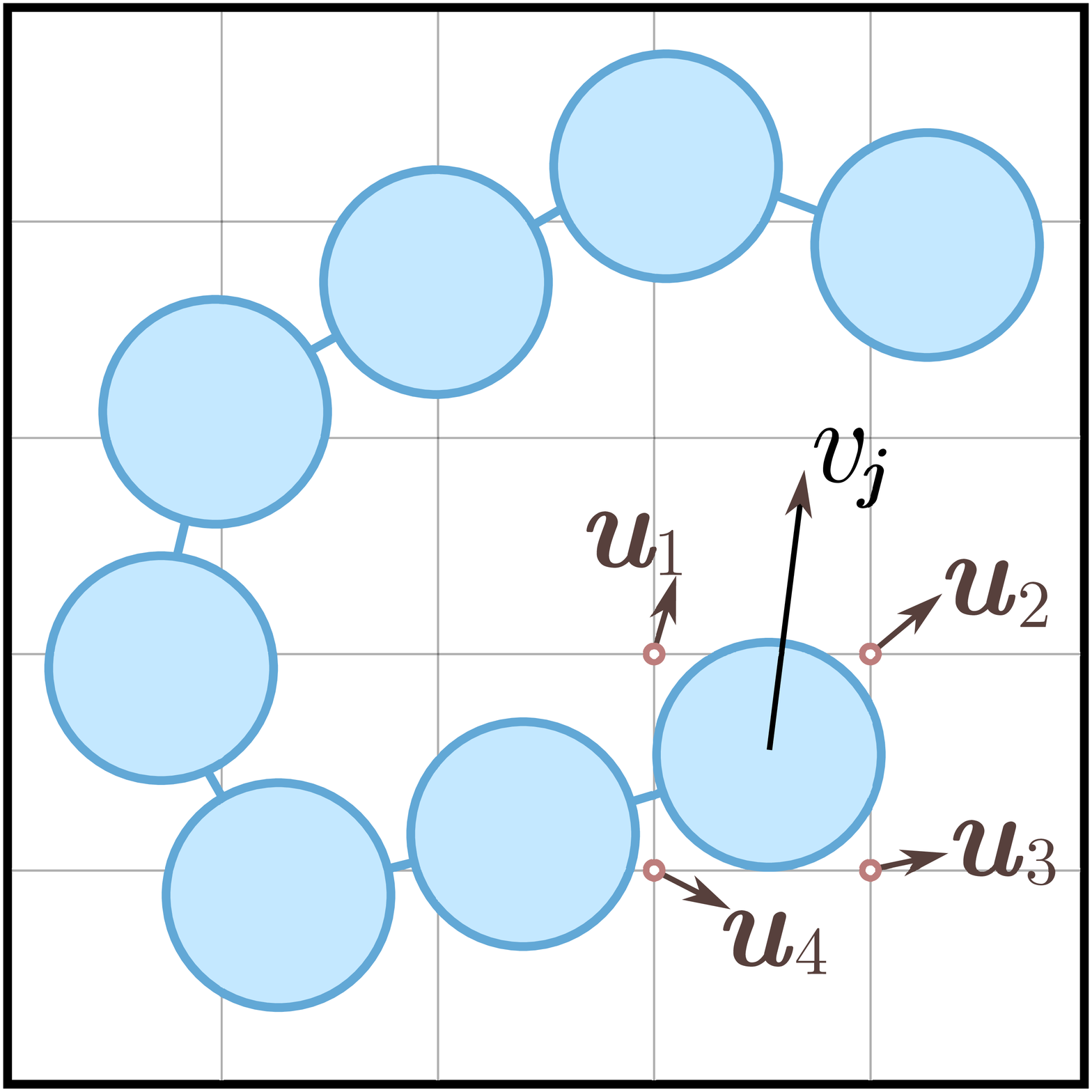}}
  \caption{(a) D3Q19 scheme with $19$ velocities connecting a chosen
    lattice site with its neighbors. (b) Schematic representation of
    the LB to MD coupling.}
\end{figure}

The LB scheme can be viewed as a version of coarse-graining of the
solvent fluid: Instead of explicit consideration of solvent molecules
and their degrees of freedom, the LB method deals with a set of
so-called populations $f_i(\vec{r}, t)$ on every lattice site
$\vec{r}$ at time $t$. The population $f_i$ is a quantity proportional
to the number of fluid particles flowing with a specific velocity,
locally at position $\vec{r}$ at time $t$. Typically, $f_i$ is
interpreted as the local mass density associated with the lattice
velocity $\vec{c}_i$. The finite set of velocities is chosen such that
in one time step neighboring lattice sites are connected. The most
popular model in three dimensions is called D3Q19. It has $19$
velocity vectors $\vec{c}_i$ (including $\vec{c}_0=0$), and is
schematically shown in Fig.~\ref{fig:d3q19}.

The hydrodynamic quantities of the fluid are found by evaluating
moments of the populations with respect to the discrete velocity set: The
mass density and the mass density flux (or momentum density) at the
lattice site $\vec{r}$ are given by

\begin{equation}
  \rho(\vec{r}) = \sum_i f_i
\end{equation}
and

\begin{equation}
  \vec{j}(\vec{r}) = 
  \rho(\vec{r}) \vec{u}(\vec{r}) = \sum_i f_i \vec{c}_i , 
\end{equation}
respectively. This also defines the local flow velocity $\vec{u}$.

The change of the populations during one LB timestep $\Delta
t_{\textrm{LB}}$ on a lattice site is summarized by the Lattice
Boltzmann equation (LBE):

\begin{equation}
f_i (\vec{r} + \vec{c}_i \Delta t_{\textrm{LB}}, t + 
\Delta t_{\textrm{LB}}) = f_i (\vec{r}, t) + \Delta_i ,
\end{equation}
where $\Delta_i$ is the so-called collision operator. The choice of
the collision operator is dictated by (i) conservation of mass and
momentum, (ii) the rate of stress relaxation in the framework of a
linearized Boltzmann equation, which determines the viscosities, (iii)
proper implementation of the thermal fluctuations, and (iv) external
forces (for example, friction forces from the MD particles, which are
the main concern of the present paper). A very general formulation is
the multiple-relaxation times (MRT) collision
operator~\cite{DdH_IG_MK_2002}, as discussed in
Ref.~\cite{BD_US_AL_2007}. After the collision step, the
post-collision populations are propagated to the neighboring sites
according to the velocity vectors $\vec{c}_i$. This is again
relatively easy to parallelize. As long as the immersed particles are
absent, or present only at a low concentration, such that their
contribution to the overall numerical effort is insignificant,
load-balancing problems do not occur. For details, see
Ref.~\cite{BD_AL_2009}.

\section{Coupling algorithm}
\label{sec:coup}

The force coupling scheme is sketched in
Fig.~\ref{fig:LB_MD_coupling}. The force $\vec{F}_j$ acting on MD
particle number $j$, which is located at position $\vec{R}_j$ and
moves with velocity $\vec{v}_j$, is given by

\begin{equation}
  \vec{F}_j = \vec{F}_j^{\textrm{cons}} 
  - \zeta \left[ \vec{v}_j - \vec{u}(\vec{R}_j) \right] 
  + \vec{F}_j^{\textrm{rand}},
  \label{eq:old_coupl}
\end{equation}
where $\vec{F}_j^{\textrm{cons}}$ is the total conservative force on
the particle, resulting from all interactions with other particles. It
has no relation to the LB fluid and we will therefore omit it from now
on in the text. The last term $\vec{F}_j^{\textrm{rand}}$ is the
random force due to thermal motion. The second term is the frictional
force due to the coupling, where $\zeta$ is the particle's friction
coefficient (here, for simplicity, assumed to be identical for all
particles), while the term in brackets is the velocity of the particle
$\vec{v}_j$ relative to the fluid velocity $\vec{u}$ at the particle's
position $\vec{R}_j$. Since the fluid velocity is only defined at the
lattice sites, one has to interpolate $\vec{u}(\vec{R}_j)$ from the
velocities at some neighboring sites. A simple interpolation scheme
involves (in 3D) eight sites $\vec{r} = \vec{r}_k$, $k = 1, \ldots,
8$, which form the cube that contains the particle:

\begin{equation} 
\vec{u}(\vec{R}_j) = \sum_{k=1}^8 \delta_k \,
\vec{u}(\vec{r}_k) , \label{eq:inter_vel} 
\end{equation}
where $\delta_k$ are weights based on the distances (in $x$, $y$ and
$z$ direction) between $\vec{R}_j$ and the sites $\vec{r}_k$. Clearly,
within one MD time step $\Delta t$, the particle's coupling to the
fluid results in a change of its momentum by an amount $\Delta
\vec{P}_j$. We here consider only a very simple integration scheme,
which evaluates the momentum change via

\begin{equation} \Delta \vec{P}_j = \Delta t \left\{ - \zeta \left[
\vec{v}_j - \vec{u}(\vec{R}_j) \right] + \vec{F}_j^{\textrm{rand}}
\right\} , \end{equation}

where we take $\vec{R}_j$, $\vec{v}_j$, and
$\vec{F}_j^{\textrm{rand}}$ as the values obtained or generated at the
beginning of the MD step.

To conserve momentum one has to consider Newton's third law and
account for the force exerted by the particle $j$ onto the
surrounding LB fluid. In other words, the total momentum of the fluid
has to be changed by the amount $- \Delta \vec{P}_j$, and locality
dictates that this amount of momentum should be distributed onto some
surrounding sites, again with some interpolation scheme.  Not
surprisingly, one should use the \textit{same} interpolation scheme as
the method that is used for the interpolation of the velocities;
analysis of the continuum version of the algorithm shows that this is
necessary in order to satisfy the fluctuation-dissipation
relation~\cite{BD_AL_2009}. Furthermore, we can transform the change
in momentum at the site $\vec{r}_k$ to a force density, which is thus
found to be

\begin{equation}
  \vec{f}_k = - \frac{1}{\Delta t \, a^3} \delta_k \Delta \vec{P}_j ,
  \label{eq:f_dens}
\end{equation}
where $a$ is the lattice spacing. Now, the problem how to apply a
force density in LB has a nice and well-defined
solution~\cite{BD_AL_2009} first reported by Guo et
al.~\cite{guo_discrete_2002}: The force density $\vec{f}_k$ gives rise
to a certain contribution $\Delta_i^\prime$ to the collision operator,
which is linear in $\vec{f}_k$ and which is adjusted in such a way
that not only the momentum transfer is correct, but also the continuum
limit obtained via a second-order Chapman-Enskog
expansion~\cite{BD_AL_2009} is just the NSE without any spurious
terms. It should be noted that $\Delta_i^\prime$ depends not only on
$\vec{f}_k$ but also on the local streaming velocity $\vec{u}$.

However, due to the fact that typically $\Delta t$ is smaller
than $\Delta t_{\textrm{LB}}$, as a result of the coarse-graining of
the LB fluid with respect to time~\cite{PA_BD_1999}, the question
remains open how to implement this precisely. In what follows, we will
discuss the method in terms of the change in momentum density
corresponding to the time $\Delta t$,

\begin{equation}
  \Delta \vec{j}_k = \vec{f}_k \, \Delta t .
\end{equation}

\begin{figure}[t]
  \centering
  \includegraphics[width=0.8\textwidth]{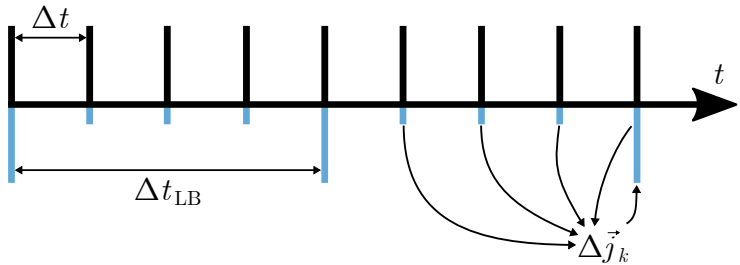}
  \caption{Schematic representation of the LB and MD timescales.}
  \label{fig:LB_MD_timescale}
\end{figure}

We denote the ratio between the two time steps by $m = \Delta
t_{\textrm{LB}} / \Delta t$. In order to keep the coupled simulations
synchronized, it is necessary that $m$ is an integer; typically we
will have $m > 1$. The situation that then arises is schematically
depicted in Fig.~\ref{fig:LB_MD_timescale}.

The coupling force $\vec{F}_j$ is computed every MD time step (black
vertical lines) and employed to integrate the equation of motion of
particle $j$. At the same time, the concomitant changes in momentum
density, $\Delta \vec{j}_k$, are made known to their respective LB
nodes $\vec{r}_k$. The simplest method (which we will call variant A
in what follows) would then accumulate these values for $m$ steps,
while keeping the populations and streaming velocities on the
surrounding sites fixed. This is indicated in
Fig.~\ref{fig:LB_MD_timescale} by black arrows. After these $m$ steps,
the collision operator (with $\Delta_i^\prime$ corresponding to the
total time $\Delta t_{\textrm{LB}}$) is applied, followed by the
streaming step (long blue vertical lines).

It is desirable to make $m$ as large as possible, in order to avoid
the CPU-intensive LB steps as much as possible. This is essential for
the computational performance of most physical applications. The more
dilute the solution is, the larger is the CPU time fraction of the LB
part. For a typical semidilute polymer solution (where the monomer
volume fraction is small) we observe that the integration of the
equations of motion of the particle-based system takes less than $5\%$
of the total time. Therefore, the ability to reduce the number of LB
collision-streaming steps speeds up the simulations considerably.

The described coupling scheme is simple and therefore computationally
efficient. It is also reasonably robust in a moderate range of
parameters. To test this, we have calculated the average temperature
of the particle-based system, evaluated by the mean kinetic energy.
The particle system consists of ten bead-spring polymer chains with
$256$ beads each, whose interactions are given by the standard
Kremer-Grest model~\cite{GG_KK_1986}: All monomers interact via a
Lennard-Jones (LJ) potential with energy parameter $\epsilon$ and
length parameter $\sigma$, where the interaction is however not cut
off at $2^{1/6} \sigma$, but rather at $2^{7/6} \sigma$ (for a reason
of this, see below). Furthermore, the monomers are connected by FENE
springs, as in the standard model (and parameters chosen as in
Ref.~\cite{GG_KK_1986}). The monomers have a mass $\mu$, and $\tau$ is
the standard Lennard-Jones time $\tau = \left( \mu \sigma^2 / \epsilon
\right)^{1/2}$. The size of the simulation box is $(20\sigma)^3$. The
LB lattice spacing is set to $a = \sigma$, and the fluid density is
$\rho = 1.0 \mu / a^3$. The desired temperature of the system is
$k_{\textrm{B}} T = 1.0 \epsilon$, and the fluid shear viscosity is
$\eta = 3 \epsilon \tau / \sigma^3$.

The algorithm was implemented as a part of the simulation package
ESPResSo++~\cite{JH_TB_OL_2013}, which employs domain-decomposition
strategies to parallelize the code via the ``MPI'' (message-passing
interface) library. In order to avoid possible load-balancing problems
as a result of too strong fluctuations in the number of particles per
processor, we distributed the system on eight processors (Xeon E5-2650
v2 CPUs with 2.60GHz and cache size of 20.48 Mb), such that each of
them had a load of $320$ particles on average.

\begin{figure}[t]
  \centering
  \includegraphics[width=1.0\textwidth]{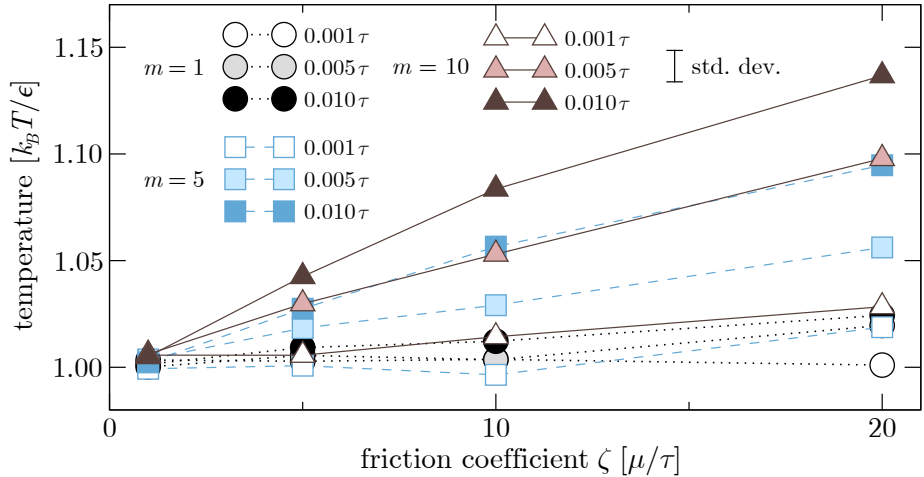} 
  \caption{Temperature of the particle-based system as a function of
    the friction coefficient $\zeta$ for variant A coupling. The data
    for time-scale contrasts of $m = 1$, $5$ and $10$ are plotted in
    black, blue and brown, respectively. The shading of the symbols
    (from open to full) corresponds to MD time steps of $0.001 \,
    \tau$, $0.005 \, \tau$ and $0.010 \, \tau$. The typical standard 
    deviation of the data is indicated as well.}
\label{fig:temp_old}
\end{figure}

By roughly estimating the polymer dimensions (assuming random-walk
conformations), and comparing them to the size of the simulation box,
one sees that the solution is well within the semidilute
regime. Furthermore, due to the attractive tail of the LJ potential,
the solvent quality is definitely poor. In fact, the temperature
$\Theta$ at which the Theta collapse of a single chain occurs has
previously been determined for this model~\cite{CP_KB_TK_MM_2006} as $k_B
\Theta / \epsilon \sim 3.3$, and for such long chains the
temperature for unmixing (below which the polymer system falls out of
solution) is only slightly smaller. Hence our simulations are clearly
in the phase-separating regime, and therefore one must assume that the
system, although having been pre-relaxed for a few million MD steps at
$\Delta t = 0.01 \, \tau$, is not yet fully in thermal equilibrium. We do not
view this as a disadvantage, since (i) we plan to apply the method in
the future precisely to such non-equilibrium systems, and (ii) we
believe that such non-equilibrium conditions put even more stringent
requirements on its robustness than running it just in equilibrium.
In this context it should be noted that processes like chain shrinkage
etc. are expected to generate significant hydrodynamic flows, beyond
the level that would occur in strict equilibrium. Due to the
thermostat, the temperature should nevertheless remain strictly
constant throughout the process, which was observed for a duration of
$4000-40000$ $\tau$ (for various MD time steps).

The result of the test is shown in Fig.~\ref{fig:temp_old}. For some
parameters, we observe significant deviations between the ``measured''
temperature and its desired input value. At fixed $m$, we find that
the deviations decrease systematically when the time step is
decreased, and therefore we interpret this behavior as a
discretization error. Furthermore, the deviations also increase
systematically with the friction constant $\zeta$. At small friction
coefficients ($\zeta \leq 5 \,\mu/\tau$), moderate MD timesteps
($\Delta t \leq 0.005 \,\tau$) and moderate time-scale contrasts ($m
\leq 5$) the deviations do not exceed $3 \%$ (cf. half-shaded squares
in Fig.~\ref{fig:temp_old}). However, outside the above mentioned
parameter region (especially for larger time-scale contrasts) the MD
system is significantly heated up. The deviations can be as large as
$15 \%$ (cf. full shaded triangles in Fig.~\ref{fig:temp_old}). It is
interesting to notice that the heat-up at fixed LB time step is the
same (cf. the systems at $m = 5$ and $\Delta t = 0.010 \tau$ {\it vs.}
$m = 10$ and $\Delta t = 0.005 \tau$).

The origin of these errors lies in the nature of the coupling: The
frictional force acting on every particle is calculated every MD step,
but for determining the reference velocity $\vec{u}(\vec{R}_j)$
variant A uses the streaming velocities on the sites $\vec{r}_k$ that
have been calculated at the last LB update at time
$t_{\textrm{LB}}^{\textrm{last}}$ (long blue vertical lines in
Fig.~\ref{fig:LB_MD_timescale}). We therefore modify variant A to a
more accurate scheme (variant B) where the MD-LB coupling is done
every MD time step, such that the time lag of the reference velocity
is substantially reduced (short and long blue vertical lines in
Fig.~\ref{fig:LB_MD_timescale}). More precisely, the scheme proceeds
as follows: At time $t = t_{\textrm{LB}}^{\textrm{last}}$ we apply the
collision operator corresponding to MRT relaxation and to thermal
noise applied to the fluid, plus $\Delta_i^\prime$ corresponding to
the total accumulated momentum transfer from the particles (on the
time scale $\Delta t_{\textrm{LB}}$). We also perform an LB streaming
step. At time $t = t_{\textrm{LB}}^{\textrm{last}} + \Delta t$, we
first apply thermal noise to the particle. After this the friction
force is calculated, using the ``old'' lattice velocities $\vec u$
obtained at $t = t_{\textrm{LB}}^{\textrm{last}}$. These forces are
used to update the particle positions and momenta by means of the MD
integrator (a velocity Verlet scheme~\cite{verlet_remark}). At the
same time, these thermal and frictional forces alter the momentum
density on the neighboring fluid sites by $\Delta \vec{j}_k$. This in
turn is used to update the lattice velocities $\vec u$ on the
surrounding sites $\vec{r}_k$. After that we can apply the same
procedure at time $t = t_{\textrm{LB}}^{\textrm{last}} + 2 \Delta t$,
where now the lattice velocities from the time $t =
t_{\textrm{LB}}^{\textrm{last}} + \Delta t$ are used to calculate the
friction force. The scheme is then continued for all together $m$
times, until at time $t = t_{\textrm{LB}}^{\textrm{last}} + \Delta
t_{\textrm{LB}}$ we again perform a full LB update. The overall
momentum conservation then holds at every single MD time step, and not
only at every LB time step as in variant A. One should thus expect
that variant B, via the reduction of the time lag between the particle
velocity and the lattice velocities at the nearby nodes, will allow
substantially larger time-scale contrasts, and thus more efficient
simulations.

\begin{figure*}[t]
  \centering
  \includegraphics[width=1.0\textwidth]{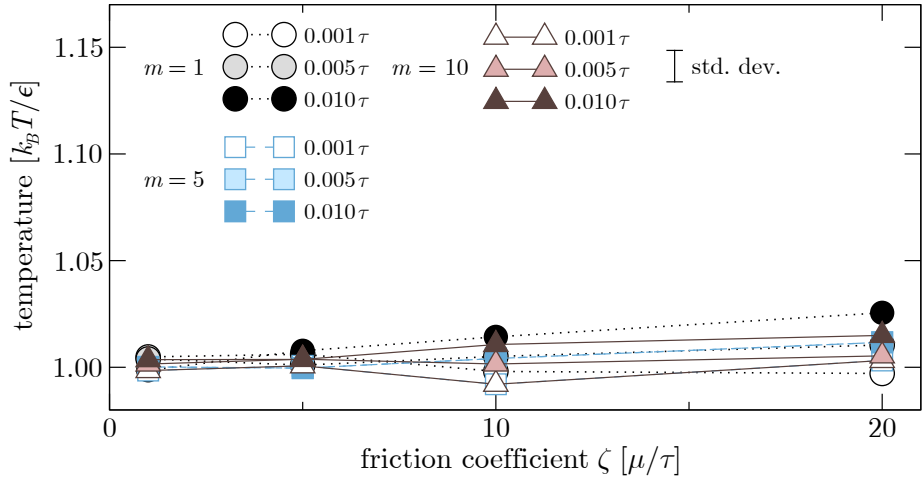}
  \caption{Same as Figure~\ref{fig:temp_old} but for variant B
    coupling.}
  \label{fig:temp_new}
\end{figure*}
        
\begin{figure*}[t]
  \centering
  \includegraphics[width=1.0\textwidth]{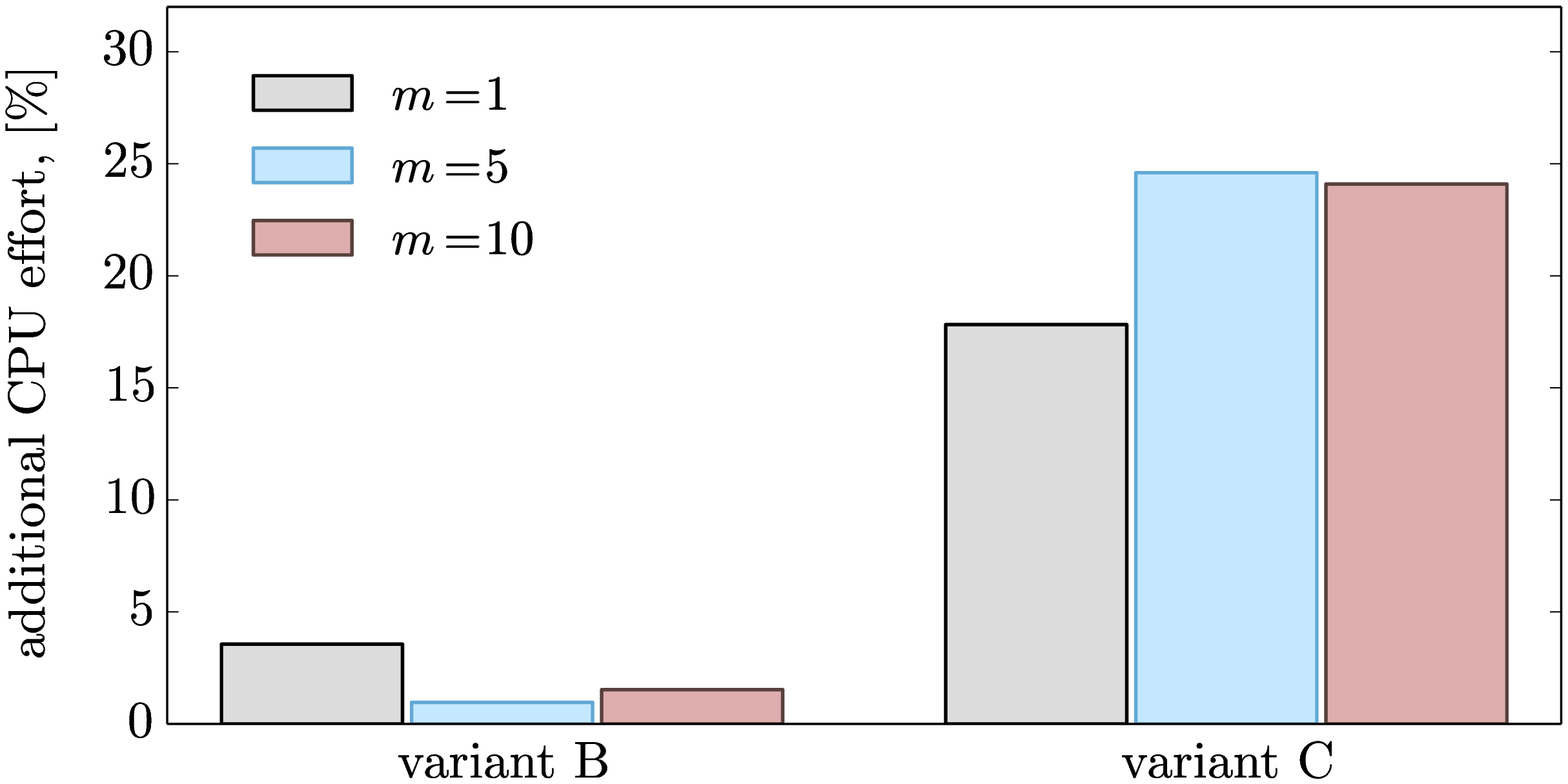}
  \caption{Additional CPU effort in variants B and C of the coupling
    with respect to the variant A for $\Delta t = 0.005 \tau$ and
    $\zeta = 20 \,\mu/\tau$.}
  \label{fig:add_cpu}
\end{figure*}

The effect that the improved coupling scheme has is enormous and can
be seen in detail in Fig.~\ref{fig:temp_new}. Even when the time-scale
contrast is as large as $m = 10$ and the MD time step takes the large
value $\Delta t = 0.01 \tau$, the difference between the observed and
the desired temperature is less than $2\%$, for all values of $\zeta$
that we have studied! It is therefore obvious that variant B is
greatly superior to variant A in the correct reproduction of the
particle kinetic energy --- and this means it has also much more
superior temperature stability. This in turn means that one can afford
a much larger time-scale contrast, and thus gain enormously in
computational efficiency. The programming effort to switch from
variant A to variant B is minimal, involving just a few lines of
code. Similarly, the additional CPU effort is, at least for dilute
systems, usually just a few percent, since only a few nodes in the
vicinity of the particles are involved (cf. Fig.~\ref{fig:add_cpu},
left panel).

\begin{figure*}[t]
  \centering
  \includegraphics[width=1.0\textwidth]{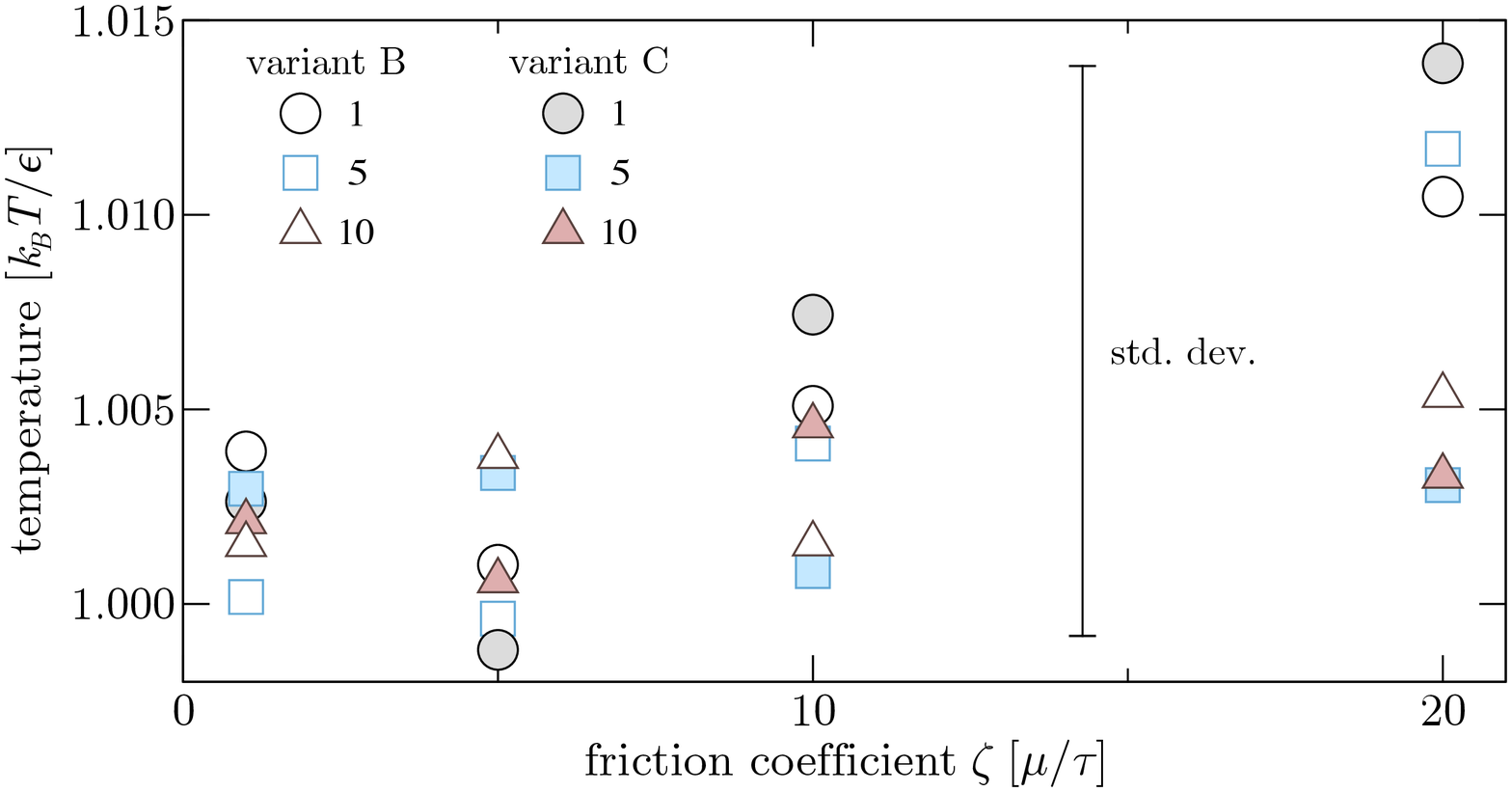}
  \caption{
    Temperature of the particle-based system as a function 
    of the friction coefficient $\zeta$ for coupling variants B and C (open
    and shaded symbols, correspondingly). 
    The data for time-scale contrasts of $m = 1$, $5$ and $10$ are plotted in
    black, blue and brown, respectively. The MD time step is fixed at 
    $0.005 \, \tau$. The typical standard deviation of the data is shown as well.
    }
  \label{fig:compare_BC}
\end{figure*}

The variant B of the coupling presented above is a simplification of a
more refined approach we will refer to as variant C. This approach
involves a full update of the \textit{populations} (rather than only
the momenta) on the lattice sites $\vec{r}_k$ at \textit{every MD time
  step}. This operation requires neither MRT relaxation nor adding
random noise to the LB fluid nor the streaming. The update is done via
the collision operator $\Delta_i^\prime$ corresponding to the momentum
transfer from the particles. Since $\Delta_i^\prime$ depends not only
on the momentum transfer, but also on the local streaming
velocity~\cite{BD_AL_2009,guo_discrete_2002}, this is genuinely
different from the approach outlined above and involves a small
nonlinear correction to the flow velocities: At the next time step the
lattice velocities are calculated from the changed populations (and
not simply from the momentum transfer) and the update procedure is
repeated again. The full LB update with streaming is done every $m$
time steps as usual.

We have also tested variant C, and the results are presented in
Figs.~\ref{fig:add_cpu} and \ref{fig:compare_BC}. Since now the LB
part of the coupling is much more involved, we observe a significant
increase of the CPU effort, which exceeds that of variant A by more
than $20 \%$ (see Fig.~\ref{fig:add_cpu}, right panel) and is hence
clearly non-negligible. On the other hand, when comparing the
temperature stability of variants B and C (Fig.~\ref{fig:compare_BC}),
we are unable to observe any improvement --- the deviations that we
observe are so small that they are below the statistical noise of the
data. Therefore we recommend the variant B as a standard coupling
scheme for MD particles interacting with an LB fluid.

Interestingly enough, one observes significant differences in the CPU
effort between the three variants even for $m = 1$, in which case the
three variants in principle coincide, meaning that they are
algorithmically equivalent. However, even for $m = 1$ they do differ
in terms of their implementation, i.~e. the way how the book-keeping
is done, and in what order the various operations are performed. The
increases in CPU effort that we observe are therefore most likely not
so much a result of an increased operation count, but rather of an
increased rate of cache misses.

\section{Conclusions}
\label{sec:conc}

The present study highlights the importance of implementing the force
coupling in such a way that it keeps the discretization errors
reasonably small, such that large time-scale contrasts (and therefore
efficient simulations) can be afforded. This goal is achieved by (i)
minimizing the time lag between the particle velocity and the
streaming velocities at the nearby nodes, and (ii) enforcing momentum
conservation at every single MD time step. At the same time, the
computational cost of the modification (in its variant B) is
negligible, since only quantities are used that are available during
the simulation anyway.

In contrast to a straightforward variant A, variant B employs the
existing coupling forces stored at the lattice sites to update the
velocities of the LB fluid every MD step. Though in variant A the
coupling forces are also calculated every MD step, they are merely
accumulated until the LB update step. Conversely, variant B proposes
to apply the coupling forces to modify the velocities of the LB fluid
immediately, thus allowing for a more accurate calculation of the
reference velocity that enters the friction force. This fine
modification has a dramatic effect on the temperature stability of the
simulation: The temperature of the MD system is kept effectively
constant in a significantly extended parameter space comprising the
friction coefficient of the coupling, the MD timestep and the
time-scale contrast between LB and MD.

We expect that the observed increase in temperature for variant A is a
precursor to a full instability of the algorithm as such, which would
occur if one would go to even more extreme parameter values. In this
context, one should note that an increase of $m$ (or the LB time step)
implies a decrease of the LB speed of sound (in MD units). Ultimately
this results in a violation of the small Mach number condition, and
the LB part becomes unphysical.

Finally, we also briefly investigated a variant C where the updates of
the nearby nodes do not only involve the streaming velocities but
rather the populations as such, resulting in an additional very small
correction of the velocities that enter the Stokes friction term. It
was shown that this leads to a temperature stability that is, within
statistical error bars, indistinguishable from that of variant B.
Since, on the other hand, this procedure results in a significant
computational overhead, it is not recommended.

We believe that it is highly plausible that the main reason for the
observed improvement is (as stated above) the reduction in time lag
between the two parts, such that momentum conservation is enforced for
every MD step instead of only every LB step. This also means that we
believe that other properties of the implementation are far less
important for this aspect of accuracy and stability --- although we
have not tested this in detail. For example, we believe that we would
obtain rather similar results if we would change the velocity / force
interpolation to a higher-order scheme that would involve more nearby
neighbors. The same comment holds for changing the MD
integrator~\cite{verlet_remark} to a more sophisticated (higher-order)
algorithm.

\section{Acknowledgments}

We thank A.~C.~Fogarty for a careful reading of the manuscript, and
her useful comments.  Stimulating discussions with U. Schiller and
A.~J.~C. Ladd are gratefully acknowledged. This work was supported by
the DFG Collaborative Research Center TRR 146 ``Multiscale Simulation
Methods for Soft Matter Systems''.

\section*{References}

\bibliography{mod_coupling}

\begin{thebibliography}{10}
\bibitem{verlet_remark}
The Verlet algorithm is implemented via (i) a momentum update with $\Delta t
  /2$, using the forces from the previous step, (ii) a position update with
  $\Delta t$, followed by a force calculation, and (iii) another momentum
  update with $\Delta t / 2$. Here the dissipative and random forces are
  calculated together with the conservative forces. For simplicity, these parts
  are treated on the level of the Euler algorithm (accurate up to $O(\Delta
  t)$) for Langevin equations.
\end{thebibliography}

\end{document}